# Hybrid Device-to-Device and Device-to-Vehicle Networks for Energy-Efficient Emergency Communications

Zhengrui Huang, *Graduate Student Member*, *IEEE*

***Abstract*—Recovering postdisaster communications has become a major challenge for search and rescue. Device-to-device (D2D) and device-to-vehicle (D2V) networks have drawn attention. However, due to the limited D2D coverage and onboard energy, establishing a hybrid D2D and D2V network is promising. In this article, we jointly establish, optimize, and fuse D2D and D2V networks to support energy-efficient emergency communications. First, we establish a D2D network by optimally dividing ground devices (GDs) into multiple clusters and identifying temporary data caching centers (TDCCs) from GDs in clusters. Accordingly, emergency data returned from GDs is cached in TDCCs. Second, given the distribution of TDCCs, unmanned aerial vehicles (UAVs) are dispatched to fetch data from TDCCs. Therefore, we establish a UAV-assisted D2V network through path planning and network configuration optimization. Specifically, optimal path planning is implemented using cascaded waypoint and motion planning and optimal network configurations are determined by multiobjective optimization. Consequently, the best tradeoff between emergency response time and energy consumption is achieved, subject to a given set of constraints on signal-to-interference-plus-noise ratios, the number of UAVs, transmit power, and energy. Simulation results show that our proposed approach outperforms benchmark schemes in terms of energy efficiency, contributing to large-scale postdisaster emergency response.**

### NOMENCLATURES

| | |
|---|---|
| BS | Base Station |
| D2D | Device-to-Device |
| D2V | Device-to-Vehicle |
| EC | Energy Consumption |
| EE | Energy Efficiency |
| ECN | Emergency Communication Network |
| EOC | Emergency Operation Center |
| ERT | Emergency Response Time |
| GD | Ground Device |
| MMP | Min-Max Problem |
| ONP | Optimal Network Partitioning |
| PF | Pareto Front |
| SINR | Signal-to-Interference-Plus-Noise Ratio |
| TDCC | Temporary Data Caching Center |
| UAV | Unmanned Aerial Vehicle |

Zhengrui Huang is with the Academy of Digital China (Fujian), Fuzhou University, Fuzhou 350108, China, and also with the Department of Geography, The Pennsylvania State University, University Park, PA 16802, USA. (email: zqh5210@psu.edu)

## I. INTRODUCTION

BUILDING emergency communication networks (ECNs) is a key challenge in rescue work, having a direct and strong influence on the safety of human beings and natural resources [1]. After natural or man-made disasters, most ground devices (GDs) cannot access dysfunctional ground base stations (BSs). Additionally, rescuers are stopped by collapsed terrains and cannot monitor real-time information. In support of emergency communications, it is crucial to establish a temporary network, allowing GDs to access emergency operation centers (EOCs). Hopefully, device-to-device (D2D) or device-to-vehicle (D2V) networks will provide reliable solutions [2]. In a D2D network, a direct link between two users is established without traveling through a BS, improving regional connectivity [3]. However, according to the sparsity of user distribution, it is impossible to establish D2D links for all users, indicating that other channels should be introduced to extend coverage. In recent years, D2V networks have gained attention, especially utilizing unmanned aerial vehicle (UAV)-aided networks to realize communication recovery [4]. However, considering the practical performance of UAVs, it is still challenging to provide ubiquitous coverage [5]. Motivated by the above, we consider that integrating D2D and D2V networks contributes to energy-efficient emergency communications. Nonetheless, there is a lack of studies on the joint establishment, optimization, and fusion of D2D and D2V networks, which is the focus of this article.

### *A. Related Works*

Only establishing a D2D or D2V network has limitations, so exploring how to integrate D2D and D2V networks to support emergency communications is worthwhile. The existing works, nonetheless, did not effectively address the mentioned issue, as presented in Table I. In [6], a long-term evolution-based D2D network was proposed to provide emergency services for GDs. Zhao et al. [1] scheduled UAVs to receive emergency data returned from ground D2D networks but did not optimize D2D transmission performance. In addition, Hu et al. [7] established a UAV-assisted ECN to improve network throughput, and Lin et al. [8] formulated a metric called prioritized delay to measure latency. Similarly, a cache-enabled UAV ECN was proposed to maximize data transmission rates [9]. However, the three works disregarded the limitation of onboard energy. By taking into account energy constraints, an energy-aware drone deployment algorithm was proposed to extend network coverage [10], and a 3-D UAV deployment and scheduling method was developed to improve channel capacity and save power [11]. In contrast, it was reasonable to jointly optimize emergency response time (ERT) and energy consumption (EC) in emergency cases [12] and [13]. Fu et al. [14] deployed UAVs as aerial BSs to collect data, supporting energy-efficient emergency communications. More recently, with the emerging of reinforcement learning, a deep $Q$-network was built to schedule UAVs in a D2V network [15], and Zhang et al. [16] designed a $Q$-learning framework for swarm emergency communications. Given the above networks, one promising solution to establishing an energy-efficient ECN is to integrate D2D and D2V networks. Compared to [1]-[16], thus, this article is the first effort to jointly establish, optimize, and fuse D2D and D2V networks to implement energy-efficient emergency communications.

TABLE I: COMPARISONS OF RELATED WORKS

| Ref. | Dim. | Number of UAVs | D2D | D2V | EE | Approach |
|---|---|---|---|---|---|---|
| [1] | 2-D | × | √ | √ | × | Convex optimization (CO) |
| [6] | 2-D | × | √ | × | × | Clustering algorithm |
| [7] | 3-D | × | × | √ | × | Successive CO |
| [8] | 3-D | × | × | √ | × | Block successive minimization |
| [9] | 3-D | × | × | √ | × | Stochastic geometry |
| [10] | 2-D | × | × | √ | √ | Dinkelbach method |
| [11] | 3-D | × | × | √ | √ | Dijkstra's algorithm |
| [14] | 3-D | × | × | √ | √ | Geometric programming method |
| [15] | 2-D | × | √ | √ | × | Deep $Q$-network |
| [16] | 3-D | × | × | √ | √ | $Q$-learning |
| ours | 3-D | √ | √ | √ | √ | CO and multiobjective optimization |

### B. Contributions

Our main contributions are summarized as follows.

• We establish a D2D network by determining the optimal network partitioning (ONP) and identifying temporary data caching centers (TDCCs). First, we propose a stepwise iterative algorithm to partition a network by jointly maximizing network modularity and minimizing outage probabilities, subject to a given set of signal-to-interference-plus-noise ratio (SINR) and transmit power constraints. Second, we determine the locations of TDCCs using multiscale feature measurements and the technique for order preference by similarity to an ideal solution (TOPSIS) model [17]. Thus, GDs with the highest connectivity will work as TDCCs to cache emergency data.

• According to the distribution of TDCCs, a D2V network is established using path planning and configuration optimization. To implement optimal path planning, we first analytically solve the optimal waypoints for a UAV by convex optimization and Taylor's approximation, which saves the transmission time for receiving emergency data cached in TDCCs. Second, given the generated waypoints, we employ optimal control to minimize movement time and design proportional-differential controllers to generate a 3-D trajectory for a UAV.

• We optimally solve D2V network configurations, including the number of UAVs and the transmit power of TDCCs, by simultaneously minimizing the maximum ERT and the total EC, which is formulated as a multiobjective optimization problem (MOP). In this article, we adopt the multiobjective evolutionary algorithm based on decomposition (MOEA/D) to address this MOP, which yields a Pareto front (PF) comprising suboptimal solutions. To select the best tradeoff point from a PF, motivated by compromise programming, we select energy efficiency (EE) as our performance indicator to assist decision-making, i.e., an optimal solution is obtained when EE achieves its maximum.

The remainder of this article is organized as follows. In Section II, we introduce the system model and the problem formulation. The hybrid D2D and D2V network is proposed in Section III. Simulation results and discussions are shown in Section IV to validate our proposed approach, and conclusions are drawn in Section V.

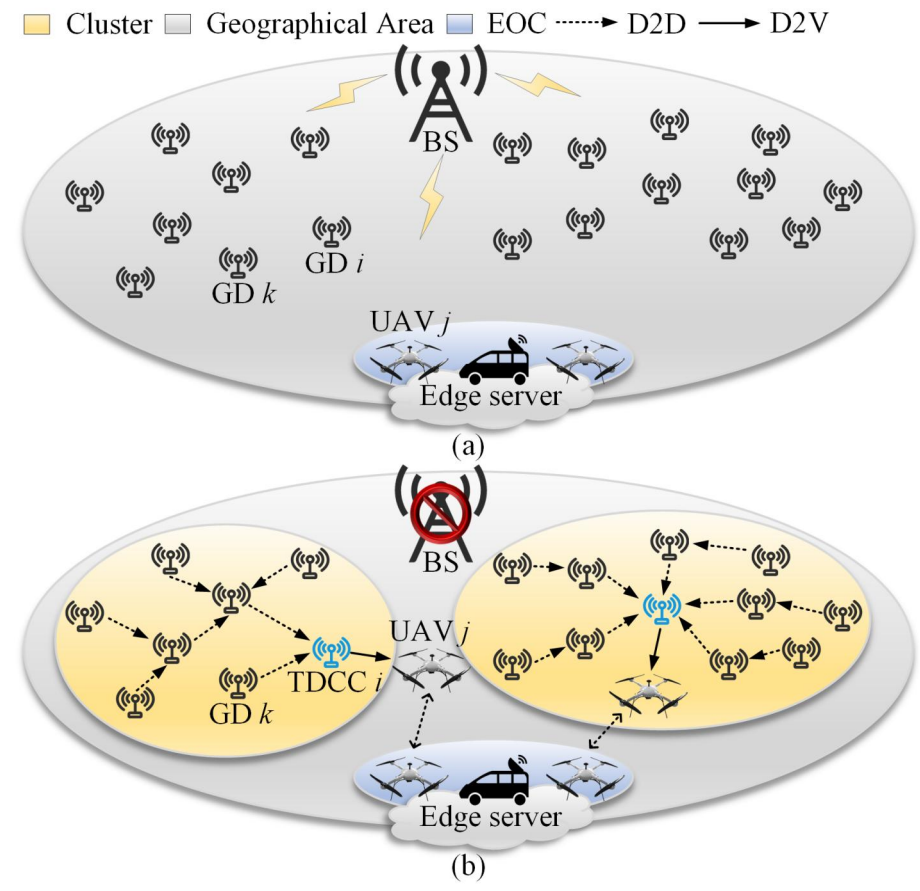

**Fig. 1.** System model. (a) Before disasters, a GD directly communicates with a BS, and UAVs are located at the EOC. (b) After disasters, this BS becomes dysfunctional. GDs will be divided into multiple clusters (each cluster has one TDCC), and the EOC will schedule UAVs to fetch data cached in TDCCs.

## II. SYSTEM MODEL AND PROBLEM FORMULATION

### A. System Model

The hybrid network includes a set $\mathcal{M}$ of $M$ four-rotary-wing UAVs and a set $\mathcal{N}$ of $N$ GDs, as shown in Fig. 1, where GDs are randomly located in a geographical area to collect emergency data and UAVs are dispatched to fetch emergency data cached in TDCCs. The locations of GD $i$ and UAV $j$ are denoted by $\left(x_i, y_i, z_i\right)$ and $\mathbf{u}_j = \left(x_j^{\text{UAV}}, y_j^{\text{UAV}}, z_j^{\text{UAV}}\right)$, respectively. In our model, we consider the following configurations: 1) The D2D network comprises multiple clusters. In each cluster, a GD with the highest connectivity functions as a TDCC, and other cluster members transmit data to the TDCC; 2) UAVs follow optimal paths to collect data cached in TDCCs in the uplink and return to the EOC to offload data and change batteries; 3) The EOC is located at an edge server that knows the locations of GDs and UAVs, and is responsible for computational tasks [1], such as implementing network partitioning and path planning; 4) The frequency-division multiple access over orthogonal channels is employed for D2D and D2V communications; 5) The chirp spread spectrum modulation scheme called LoRa is applied to support long-range and low-power emergency communications [4]; 6) The carrier frequency and bandwidth are 433 MHz and 1 MHz, respectively, the spreading factor and coding rate are set to 5.0 and 1.0, respectively, and the upper bound of transmit power is denoted by $P_{\max}$ [18].

*1) User Distribution Model:* The user distribution is modeled as a graph, denoted by **G**, in which GD $i$ has a certain chance of communicating with GD $k$ [19]. In this article, a homogeneous Poisson point process (HPPP) is utilized to generate a random graph **G** over a digital elevation model (DEM), and the degree distribution of **G** obeys a Poisson distribution.

*2) Channel Model:* To characterize D2D and D2V channels, the following two channel models are introduced to reflect the random attenuation caused by distances and barriers. The D2D path loss is denoted by a multipath and shadow fading-based model [20]

$$L_{i,k}^{\text{D2D}} = 10\log_{10}\left(d_{i,k}^2\left(K\alpha_{i,k}\beta_{i,k}\right)^{-1}\right), \tag{1}$$

where $K$ is a system constant derived from the carrier frequency, $\alpha \sim \exp(1)$ denotes the multipath fading gain, $\beta \sim \mathcal{N}\left(0, \sigma^2\right)$ denotes the shadow fading gain, $d_{i,k}$ is the distance between GD $i$ and GD $k$, and $\sigma$ is the standard deviation.

In contrast, the D2V path loss comprises line-of-sight (LoS) and non-LoS (NLoS) cases, whose probabilistic model is [21]

$$\begin{cases} \mathbb{P}_{\text{LoS}}^{i,j} = \left\{1 + a\exp\left[-b\left(\varphi_{i,j} - a\right)\right]\right\}^{-1}, & \text{LoS link,} \\ \mathbb{P}_{\text{NLoS}}^{i,j} = 1 - \mathbb{P}_{\text{LoS}}^{i,j}, & \text{NLoS link,} \end{cases} \tag{2}$$

where $a$ and $b$ are constants, and $\varphi_{i,j}$ is the elevation angle. Therefore, the average D2V path loss is expressed as

$$L_{i,j}^{\text{D2V}} = \mathbb{P}_{\text{LoS}}^{i,j}\left(10\log_{10}\frac{\left(4\pi d_{i,j}\right)^2}{G_l\lambda^2} + \eta_{\text{LoS}}\right) + \mathbb{P}_{\text{NLoS}}^{i,j}\left(10\log_{10}\frac{\left(4\pi d_{i,j}\right)^2}{G_l\lambda^2} + \eta_{\text{NLoS}}\right), \tag{3}$$

where $d_{i,j}$ is the distance between GD $i$ and UAV $j$, $G_l$ is the antenna gain, $\lambda$ is the wavelength, and $\eta_{\text{LoS}}$ and $\eta_{\text{NLoS}}$ are the additional path losses for LoS and NLoS, respectively.

*3) Environmental Model:* Environmental constraints denote a mathematical space restricting the topology of D2D networks and the motion space of UAVs. To guarantee that each UAV can fly along a collision-free path, the terrestrial constraint is defined as $\left\|\mathbf{u}_j, \mathbf{d}_l\right\| \geq d_0, \forall l \in \mathcal{D}$, where $\mathcal{D}$ is the set of points in a DEM, $d_0 \sim \mathcal{N}\left(\mu_0, \sigma_0^2\right)$, and $\mathbf{d}_l = \left(x_l, y_l, z_l\right)$.

Additionally, a UAV can be affected by winds whose impact is expressed as

$$\mathbf{f} = \left[\left\|\mathbf{d}_m, \mathbf{d}_n\right\|, \left\|\mathbf{d}_m, \mathbf{d}_n\right\|\cot(\Omega), \left\|\mathbf{d}_m, \mathbf{d}_n\right\|\sin(\varphi)\cot(\Omega)\right], \forall m, n \in \mathcal{D}, \tag{4}$$

where $\Omega$ is the wind speed angle. To make **f** parallel to a DEM, **f** is rotated with $\mathbf{o} = \mathbf{f} \times \mathbf{f}' \times \mathbf{z}_e$ by **R**, where $\mathbf{f}'$ is the wind speed vector next to **f**, $\mathbf{z}_e$ is the unit vector in the earth frame, and **R** is the rotation matrix [22].

## *B. Problem Formulation*

Since our goal is to establish a hybrid D2D and D2V network, the optimization problem comprises the following two parts.

*1) Problem 1:* The D2D network is modeled as a weighted directed graph, denoted by $\mathbf{G} = \left\{\mathbf{L}, \mathbf{W}\right\}$, where **L** is the adjacent matrix whose entry is $l_{i,k}$ and **W** is the weight matrix whose entry is $w_{i,k}$. Specifically, $l_{i,k}$ stands for a variable that describes the connectivity between GD $i$ and GD $k$

$$\begin{cases} l_{i,k} = 1, & \text{if } \mathbb{P}_{\text{out}}^{i,k} \leq \varepsilon, \\ l_{i,k} = 0, & \text{otherwise,} \end{cases} \tag{5}$$

where $\varepsilon$ is the threshold of $\mathbb{P}_{\text{out}}^{i,k}$, which meets $P_0 = \mathbb{P}_{\text{out}}^{i,k}\left(\varepsilon\right)^{-1}$, and the outage probability $\mathbb{P}_{\text{out}}^{i,k}$ is defined as

$$\mathbb{P}_{\text{out}}^{i,k} = \Pr\left(P_{i,k}K\alpha_{i,k}\beta_{i,k}d_{i,k}^{-2} \leq P_0\right) \overset{\alpha \sim \exp(1)}{=} 1 - \exp\left(-P_0 d_{i,k}^2\left(KP_{i,k}\beta_{i,k}\right)^{-1}\right), \tag{6}$$

where $P$ is the transmit power, $P_0$ is the minimum received power, and $w_{i,k}$ is expressed as

$$w_{i,k} = \mathbb{N}\left(l_{i,k}\mu_{i,k}R_{i,k}^{-1}\right), \tag{7}$$

where $\mu_{i,k}$ is the packet size and $R_{i,k}$ is the channel capacity.

Given (5)-(7), we attempt to optimally divide **G** into multiple clusters. To jointly model and optimize this process, a benefit function called modularity $Q$ is employed to measure whether a division of **G** is acceptable

$$\text{(P1):} \quad \max_{\mathbf{P},\mathbf{L}} \; \left(2L\right)^{-1}\sum_{i\in\mathcal{N}}\sum_{k\in\mathcal{N}}\left[\left(l_{i,k} - L^{-1}l_i^{\text{in}}l_k^{\text{out}}\right)\mathbb{S}\left(c_i, c_k\right)\right] \tag{8}$$

$$\text{s.t.} \quad (1), (5)-(7), \forall i, k \in \mathcal{N}, \tag{9}$$

where **P** is the power matrix whose entry is $P_{i,k}$, $l_i^{\text{in}} = \sum_{k\in\mathcal{N}} l_{k,i}$ and $l_k^{\text{out}} = \sum_{i\in\mathcal{N}} l_{k,i}$ are in-degrees and out-degrees, respectively, $L = \sum_{i\in\mathcal{N}}\sum_{k\in\mathcal{N}} l_{i,k}$, $c$ is the cluster label, and $\mathbb{S}\left(c_i, c_k\right) = 1$ if $c_i = c_k$; otherwise, $\mathbb{S}\left(c_i, c_k\right) = 0$.

*2) Problem 2:* Given the distribution of TDCCs derived from (P1), UAVs are dispatched to fetch emergency data cached in TDCCs. Here, we aim to simultaneously minimize ERT and EC. Since UAVs and TDCCs simultaneously operate, the practical

ERT depends on the maximum ERT of different system units, which is equivalent to a min-max problem (MMP), whereas EC is derived from both UAVs and TDCCs. Motivated by this, our optimization problem is defined as the following MOP

$$\text{(P2)}: \min_{\mathcal{M},\{\mathbf{u}_j\},\{P_{i,j}\}} \quad \max\left\{\{T_j\},\{T_i\}\right\} \tag{10}$$

$$\min_{\mathcal{M},\{\mathbf{u}_j\},\{P_{i,j}\}} \quad \sum_{j\in\mathcal{M}} E_j + \sum_{i\in\mathcal{T}} E_i \tag{11}$$

$$\text{s.t.} \quad (1),(3), \sum_{j\in\mathcal{M}} \chi_j = |\mathcal{M}|, E_j \le E_{\max}, \forall j \in \mathcal{M}, \tag{12}$$

where $\chi_j \in \{0,1\}$ is a binary variable, $\chi_j = 1$ if the EOC has successfully received UAV *j*'s data; otherwise, the EOC will reschedule UAV *j* to fetch data from its assigned TDCC, $\mathcal{T}$ is the set of TDCCs, $T_j$ and $T_i$ are the ERT of UAV *j* and TDCC *i*, respectively, $E_j$ and $E_i$ are the EC of UAV *j* and TDCC *i*, respectively, and $E_{\max}$ is the upper bound of UAV *j*'s energy, which guarantees the flight continuity of UAV *j*.

For simplicity, let $T_j = t_m^j + \sum_{i\in\mathcal{S}_j} t_d^{i,j}$, where $\sum_{j\in\mathcal{M}} |\mathcal{S}_j| = |\mathcal{T}|$, $\mathcal{S}_j$ is the set of TDCCs that are assigned to UAV *j*, $t_m$ is the aerial movement time, $t_d$ is the data transmission time, and $T_i$ is expressed as [23]

$$T_i = t_d^i + t_p^i + t_q^i + t_o^i, \tag{13}$$

where $t_p$ is the propagation time, $t_q$ is the queuing delay, and $t_o$ is the operation time depending on the CPU cycles per unit time (e.g., $1\times10^{10}$ CPU cycles/s).

Using (13), the EC of UAV *j* and TDCC *i* is expressed as $E_j = P_m^j t_m^j + \max\left\{P_{\text{hover}} t_d^{i,j}\right\} + \sum_{i\in\mathcal{S}_j}\left(P_{i,j} 10^{-\left(L_{i,j}^{\text{D2V}}/10\right)} t_d^{i,j}\right)$ and $E_i = P_{i,j} t_d^{i,j}$, respectively, where $P_m$ is the aerial movement power and $P_{\text{hover}}$ is the hovering power [24]. According to the kinematics, $P_{\text{hover}}$ is a constant, and $P_m$ is decomposed into

$$P_m = P_v + P_h, \tag{14}$$

where $P_v$ is the vertical movement power and $P_h$ is the horizontal movement power [25]

$$\begin{cases} P_v = \begin{cases} \dfrac{mg}{2} v_v + \dfrac{mg}{2}\sqrt{v_v^2 + \dfrac{2mg}{\rho\pi r^2}}, & \text{climbing,} \\ \dfrac{mg}{2} v_v - \dfrac{mg}{2}\sqrt{v_v^2 + \dfrac{2mg}{\rho\pi r^2}}, & \text{descending,} \end{cases} \\ P_h = \dfrac{1}{2}\rho c_d A_e v_h^3 + \dfrac{\pi}{4} N_b c_b \rho c_d \omega^3 r_d^4 \left[1 + 3\left(\dfrac{v_h}{\omega r_d}\right)^2\right] + \omega r m g \times \ell, \end{cases} \tag{15}$$

where $v_v$ is the vertical velocity, $v_h$ is the horizontal velocity, $c_d$ is the drag coefficient, $c_b$ is the blade chord, $N_b$ is the number of blades, and $\ell$ is solved by

$$2\pi\rho\omega^2 r_d^4 \ell \sqrt{\frac{v_h^2}{\omega^2 r_d^2} + \ell^2} = mg, \tag{16}$$

where *m* is the mass, *g* is the gravity acceleration, $\rho$ is the air density, $\omega$ is the angular velocity, and $r_d$ is the rotor disk radius.

## III. Hybrid D2D and D2V Network

### A. Optimal Solution to (P1)

Establishing a D2D network has been demonstrated to be a polynomial problem (PP). If we aim to optimize this PP, it will become a nondeterministic PP, equivalent to a clique-cutting problem. Since it is difficult to directly solve (P1), we propose a stepwise iterative algorithm that decomposes (P1) into two subproblems: 1) Adjacent matrix **L** is optimized subject to a given power matrix; 2) Power matrix **P** is optimized subject to a given adjacent matrix. Thus, the ONP is obtained by iteratively optimizing **P** and **L**, and the locations of TDCCs are identified using multiscale feature measurements.

*1) Optimal Network Partitioning:* Following (P1), it is clear that maximizing *Q* is equivalent to minimizing the number of clusters. Since **G** is a weighted directed graph, (P1) is expanded to the first subproblem

$$\text{(P1-a)}: \max_{\mathbf{L}} \quad W^{-1}\sum_{i\in\mathcal{N}}\sum_{k\in\mathcal{N}}\left[\left(w_{i,k} - W^{-1} w_i^{\text{in}} w_k^{\text{out}}\right)\mathbb{S}\left(c_i, c_k\right)\right], \tag{17}$$

where $W = \sum_{i\in\mathcal{N}}\sum_{k\in\mathcal{N}} w_{i,k}$, $w_i^{\text{in}} = \sum_{k\in\mathcal{N}} w_{k,i}$, and $w_k^{\text{out}} = \sum_{i\in\mathcal{N}} w_{k,i}$.

*Proposition 1:* Maximizing (17) is equivalent to optimizing the following label vector

$$\mathbf{s} = \left[s_1, \ldots, s_i\right], \forall i \in \mathcal{N}, \tag{18}$$

where $s_i \in \{1,-1\}$, which meets $\sum_{i\in\mathcal{N}} s_i^2 = |\mathcal{N}|$.

*Proof:* First, we rewrite (17) in the following compact form

$$\max_{\mathbf{s}} \quad \left(4W\right)^{-1}\mathbf{s}^T\left(\mathbf{B} + \mathbf{B}^T\right)\mathbf{s}, \tag{19}$$

where the entry of **B** is $w_{i,k} - W^{-1} w_i^{\text{in}} w_k^{\text{out}}$ and $\mathbb{S}\left(c_i, c_k\right)$ is denoted by $\frac{1}{2}\left(s_i s_k + 1\right)$.

Second, the singular value decomposition is employed to decompose $\mathbf{B} + \mathbf{B}^T$ as $\mathbf{U}\mathbf{S}\mathbf{V}^T$, where **S** is the singular value matrix. Since $\mathbf{B} + \mathbf{B}^T$ is a square matrix, (19) is equivalent to

$$\max_{\mathbf{s}} \quad \left(4W\right)^{-1}\mathbf{s}^T\mathbf{U}'\mathbf{S}'\mathbf{V}'\mathbf{s}, \tag{20}$$

where $\mathbf{S}'$ comprises the eigenvalues of $\mathbf{B} + \mathbf{B}^T$, denoted by $\{b_i\}$, $b_i$ is placed along the diagonal of $\mathbf{S}'$, and $\mathbf{U}' = \mathbf{V}'$ comprises the eigenvectors of $\mathbf{B} + \mathbf{B}^T$, denoted by $\{\mathbf{v}_i\}$. Using (18), (20) is simplified to

$$\max_{\mathbf{s}} \quad \left(4W\right)^{-1}\sum_{i\in\mathcal{N}} b_i\left(\mathbf{v}_i^T\mathbf{s}\right)^2, \tag{21}$$

where $\mathbf{v}_i^T\mathbf{s} = 0$ if $\mathbf{v}_i^T$ is perpendicular to **s**. To optimize (21), we maximize $\mathbf{v}_i^T\mathbf{s}$ by paralleling $\mathbf{v}_i^T$ to **s** as close as possible, i.e., by sorting $\{b_i\}$ in descending order $\left(b_1 \ge \ldots \ge b_i \ge \ldots \ge b_N\right)$, $s_j = 1$ if $v_{1,j} > 0, \forall v_{1,j} \in \mathbf{v}_1^T$; otherwise, $s_j = -1$, which proves the proposition.

Following Proposition 1, a bisection scheme (Algorithm 1) is proposed to divide **G** into multiple clusters, as shown in Fig. 2. The main steps are listed as follows.

1) Initially, all GDs belong to the same cluster.
2) Label vector **s** is updated using Proposition 1.
3) Capacity threshold $C_{\max}$ is employed to constrain the maximum number of GDs in each cluster.
4) $\Delta Q = (\mathbf{s})^T \left(\mathbf{B}_\mathbf{g} + \mathbf{B}_\mathbf{g}^T\right)(\mathbf{s})(4W)^{-1}$, where **g** is the subgraph of **G** and $\mathbf{B}_\mathbf{g}(i,k) = \mathbf{B}(i,k) - \mathbb{S}(c_i, c_k)\sum_{l\in\mathbf{g}} \mathbf{B}(i,l)$.

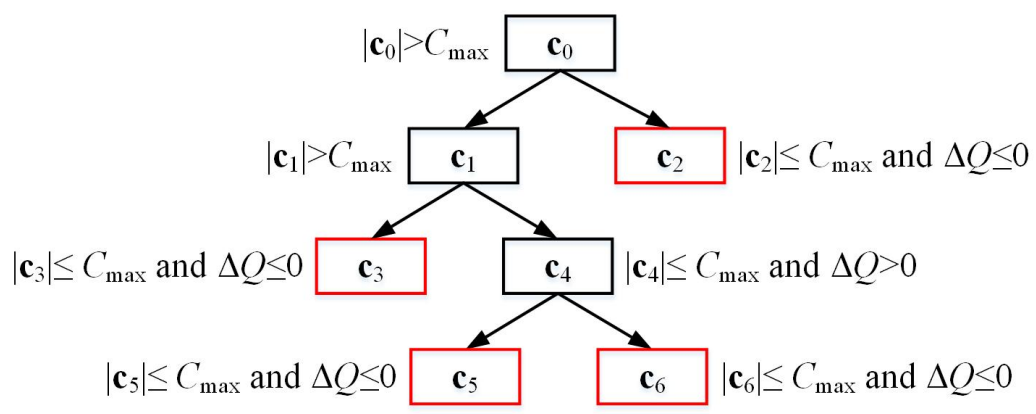

**Fig. 2.** Diagram of Algorithm 1. $\mathbf{c}_0$ is divided into $\mathbf{c}_2$, $\mathbf{c}_3$, $\mathbf{c}_5$, and $\mathbf{c}_6$.

---

**Algorithm 1** Algorithm for Dividing **G**

---

1: **Input:** Weighted directed graph **G**
2: **Output:** Set of clusters $\mathcal{C}$
3: **Initialize:** $\mathcal{C} \leftarrow \{\mathbf{c}_0\}$
4: **for** $\mathbf{c}_i \in \mathcal{C}$ **do**
5: Calculate $Q(\mathbf{c}_i)$ using a unit label vector $\mathbf{s} = [1,...,1]$
6: Calculate $\mathbf{S}'$ and update $b_1$ and $\mathbf{v}_1$ using (20)-(21)
7: **for** $v_{1,j} \in \mathbf{v}_1$ **do**
8: **if** $v_{1,j} \le 0$ **then**
9: $s_j \leftarrow -1$
10: **end if**
11: **end for**
12: Calculate $\Delta Q$ and generate $\mathbf{c}_i'$ and $\mathbf{c}_i \setminus \mathbf{c}_i'$ using **s**
13: **if** $|\mathbf{c}_i| > C_{\max}$ **then**
14: Remove $\mathbf{c}_i$ from $\mathcal{C}$ and add $\mathbf{c}_i'$ and $\mathbf{c}_i \setminus \mathbf{c}_i'$ to $\mathcal{C}$
15: **else**
16: **if** $\Delta Q > 0$ **then**
17: Remove $\mathbf{c}_i$ from $\mathcal{C}$ and add $\mathbf{c}_i'$ and $\mathbf{c}_i \setminus \mathbf{c}_i'$ to $\mathcal{C}$
18: **end if**
19: **end if**
20: **end for**
21: return $\mathcal{C}$

---

Given the updated adjacent matrix **L**, power matrix **P** is optimized by maximizing $Q$. Ideally, the optimal $Q$ is obtained when GDs intercommunicate. Based on this, (P1) is converted into the second subproblem

$$\text{(P1-b):} \quad \min_{\mathbf{P}} \quad \sum_{i\in\mathcal{N}}\sum_{k\in\mathcal{N}} 1 - \exp\left(-P_0 d_{i,k}^2 \left(K P_{i,k} \beta_{i,k}\right)^{-1}\right) \tag{22}$$

*Proposition 2:* Optimizing (22) is equivalent to minimizing variable $P$.

*Proof:* First, variable $P$ is regarded as an independent variable, and its first derivative is denoted by

$$\frac{\partial \mathbb{P}_{\text{out}}^{i,k}}{\partial P_{i,k}} = -P_0 d_{i,k}^2 \left(K \beta_{i,k}\right)^{-1} \exp\left(-P_0 d_{i,k}^2 \left(K P_{i,k} \beta_{i,k}\right)^{-1}\right) P_{i,k}^{-2}. \tag{23}$$

Since (23) is smaller than zero, minimizing (22) is equivalent to maximizing variable $P$

$$\max_{P_{i,k}} \quad P_{i,k} \tag{24}$$

$$\text{s.t.} \quad \gamma \le \frac{P_{i,k} 10^{-\left(L_{i,k}^{\text{D2D}}/10\right)}}{N_0 + \sum_{l\in\mathcal{L}\setminus i} P_{l,k} 10^{-\left(L_{l,k}^{\text{D2D}}/10\right)}}, \; P_0 10^{\left(L_{i,k}^{\text{D2D}}/10\right)} \le P_{i,k} \le P_{\max}, \tag{25}$$

where $\mathcal{L}$ is the set of GDs connected with GD $k$, $N_0$ is the noise power, $P_0 = \gamma\left(N_0 + \sum_{l\in\mathcal{L}\setminus i} P_{l,k} 10^{-\left(L_{l,k}^{\text{D2D}}/10\right)}\right)$, and the optimal SINR of GD $i$ is $P_{\max} 10^{-\left(L_{i,k}^{\text{D2D}}/10\right)}\left(N_0 + \sum_{l\in\mathcal{L}\setminus i} P_0\right)^{-1}$.

However, the optimal SINR of GD $i$ cannot be obtained due to self-interference, i.e., other GDs also receive the interference of GD $i$. To jointly maximize SINR $i$ and minimize interference power $i$, $P_{i,k}^*$ depends on the boundary constraints in (25). Fig. 3 compares the relationship between the reciprocal of the SINR and interference power. It is observed that $P_{i,k}$ converges to its lower bound, which proves the proposition.

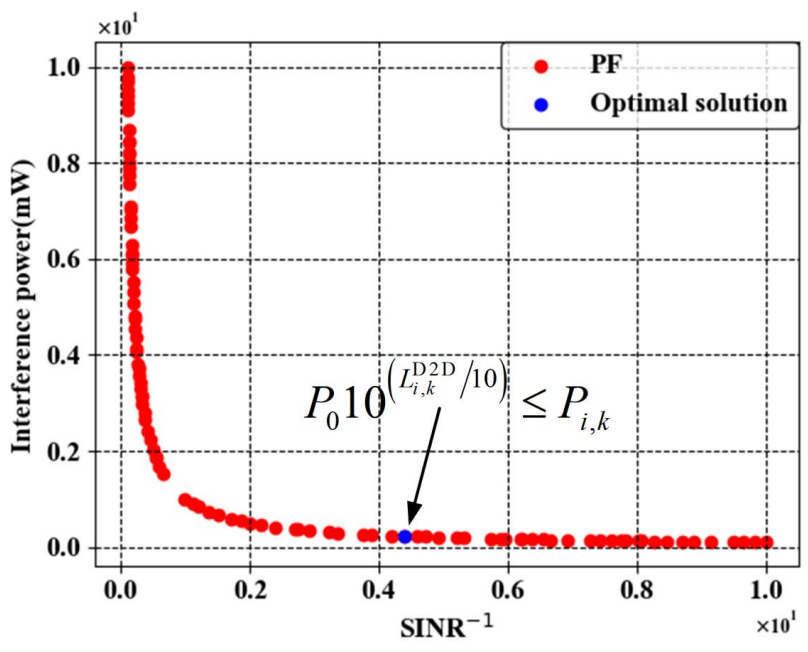

**Fig. 3.** Convergence of $P_{i,k}$.

By iteratively solving (P1-a) and (P1-b), the optimal number of clusters is obtained, as shown in Algorithm 3, including two processes: 1) The modularity is updated by optimizing adjacent matrix **L**; 2) Power matrix **P** is updated by minimizing outage probabilities, as shown in Algorithm 2.

---

**Algorithm 2** Algorithm for Updating **G**

---

1: **Input:** Weighted directed graph **G**, number of GDs $N$, and power matrix **P**
2: **Output:** Updated **G**
3: **Initialize:** Number of iterations $\leftarrow N$
4: **for** $i = 1,...,N$ **do**
5: **for** $k = 1,...,N$ **do**
6: Update $P_{i,k}$ using (1) and (25)
7: Update $\mathbb{P}_{\text{out}}^{i,k}$ using (6)
8: Update $l_{i,k}$ and $w_{i,k}$ using (5) and (7)
9: **end for**
10: **end for**
11: return **G**

---

**Algorithm 3** Algorithm for Optimal Network Partitioning

1: **Input:** Weighted directed graph **G**, number of iterations *N*, and accuracy threshold $\ell$
2: **Output:** Set of clusters $\mathcal{C}$
3: **Initialize:** Initial modularity $\leftarrow Q_0$ and initial set of clusters $\leftarrow \mathcal{C}_0$
4: **for** $i = 1,...,N$ **do**
5: Update **G** using **Algorithm 2**
6: Update $\mathcal{C}_i$ using **Algorithm 1** and calculate $Q_i$
7: **if** $|Q_0 - Q_i| \le \ell$ or $i = N$ **then**
8: return $\mathcal{C}_i$
9: **end if**
10: $Q_0 \leftarrow Q_i$
11: **end for**
12: return $\mathcal{C}_i$

The computational complexities of Algorithms 1-2 are denoted by $\mathcal{O}(|\mathcal{C}||\mathbf{v}_1|)$ and $\mathcal{O}(N^2)$, respectively, where $|\mathcal{C}|$ is the number of clusters and $|\mathbf{v}_1|$ is derived from Proposition 1. Algorithm 3 integrates Algorithms 1-2, so its computational complexity depends on $\mathcal{O}(|\mathcal{C}||\mathbf{v}_1|N^3)$.

*2) Optimal Locations of TDCCs:* After network partitioning, the locations of TDCCs are identified using multiscale feature measurements. In this article, five centrality measurements are selected to approximate the connectivity of each GD, including in-degree centrality (IC), out-degree centrality (OC), closeness centrality (CC), betweenness centrality (BC), and eigenvector centrality (EC). Specifically, IC and OC measure the popularity of GD *i*

$$\begin{cases} \mathbb{F}_{\mathrm{IC}}^{i} = (|\mathcal{N}|-1)^{-1} \sum\limits_{k\in\mathcal{N}} l_{k,i}. \\ \mathbb{F}_{\mathrm{OC}}^{i} = (|\mathcal{N}|-1)^{-1} \sum\limits_{k\in\mathcal{N}} l_{i,k}. \end{cases} \tag{26}$$

Similarly, CC estimates the proximity of GD *i*

$$\mathbb{F}_{\mathrm{CC}}^{i} = (|\mathcal{N}|-1)\left(\sum_{k\in\mathcal{N}} u_{i,k}\right)^{-1}, \tag{27}$$

where $u_{i,k}$ describing the shortest path (SP) between GD *i* and GD *k* is determined by an SP routing protocol [26].

The ability of information dissemination is defined as

$$\mathbb{F}_{\mathrm{BC}}^{i} = \sum_{k\in\mathcal{N}} \sum_{l\in\mathcal{N}} u_{k,l}(i) u_{k,l}^{-1}, \tag{28}$$

where $u_{k,l}(i)$ is the number of SPs passing through GD *i*.

The comprehensiveness of GD *i* is given by EC [27]

$$\mathbb{F}_{\mathrm{EC}}^{i} = w^{*} \sum_{k\in\mathcal{N}} (l_{i,k} e_k), \tag{29}$$

where $w^*$ is the reciprocal of the maximum eigenvalue of **W** and $e_k$ is the entry of the eigenvector of $w^*$.

To quantify the connectivity of GD *i*, (26)-(29) are input into the TOPSIS model. Fig. 4 shows five alternatives as the results based on two criteria. Point C is the closest to an ideal point, but point D is the farthest from an anti-ideal point. Therefore, the TOPSIS model measures the priority of a point, and the main steps are listed as follows.

1) An evaluation matrix $\mathbf{X}_{|\mathbf{c}|\times 5}$ for cluster **c** is created, and $\mathbf{X}_{|\mathbf{c}|\times 5}$ is normalized to eliminate the scalar effect.

2) The vector of measurement weights, denoted by $\mathbf{w}_{1\times 5}$, is given by an entropy weight method, and the normalized weight matrix **Y** is denoted by $\mathbf{X}_{|\mathbf{c}|\times 5}\mathbf{w}_{1\times 5}^{T}$, whose entry is *y*.

3) The sets of positive and negative solutions are denoted by $\mathbf{S}^{+} = \{\max\{y_{c,1}\},...,\max\{y_{c,5}\}\}, \forall c \in \mathbf{c}$ and $\mathbf{S}^{-} = \{\min\{y_{c,1}\},...,\min\{y_{c,5}\}\}, \forall c \in \mathbf{c}$, respectively.

4) $D_c^{+} = \sqrt{\sum_{n=1}^{5}(s_n^{+} - y_{c,n})^2}, \forall s_n^{+} \in \mathbf{S}^{+}$, and $D_c^{-} = \sqrt{\sum_{n=1}^{5}(s_n^{-} - y_{c,n})^2}, \forall s_n^{-} \in \mathbf{S}^{-}$.

5) The connectivity of GD *c* is given by $D_c^{-}(D_c^{-} + D_c^{+})^{-1}$.

Consequently, a GD with the highest connectivity in a cluster will function as a TDCC, and its cluster members will transfer data to the TDCC.

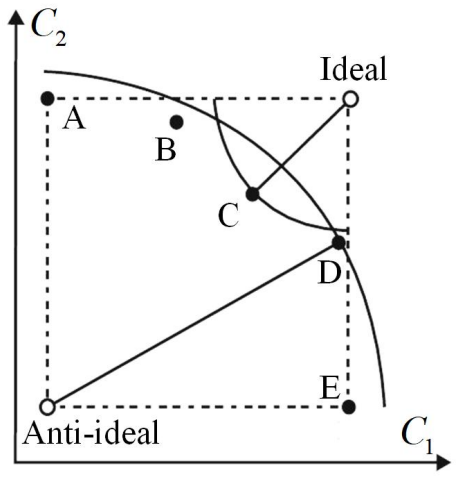

**Fig. 4.** Diagram of the TOPSIS model.

### *B. Optimal Solution to (P2)*

In this section, a UAV-assisted D2V network is proposed to receive emergency data returned from TDCCs. Here, we update $\{\mathbf{u}_j\}$ to generate optimal paths and optimize $\mathcal{M}$ and $\{P_{i,j}\}$ to obtain optimal network configurations.

*1) Optimal Path Planning:* As discussed in [18], a UAV path includes a sequence of waypoints and multiple trajectories that connect two adjacent waypoints. From (10), first, we derive the optimal waypoints by minimizing $t_d$. Second, we minimize $t_m$ to generate the expected 3-D trajectories.

To shorten the transmission time $t_d$, the waypoint planning problem is defined as

$$\text{(P2-a):} \quad \max_{P_{i,j}, L_{i,j}^{\mathrm{D2V}}} \frac{P_{i,j} 10^{-\left(L_{i,j}^{\mathrm{D2V}}/10\right)}}{N_0 + \sum\limits_{k\in\mathcal{S}_j\setminus i} P_{k,j} 10^{-\left(L_{k,j}^{\mathrm{D2V}}/10\right)}}. \tag{30}$$

To solve (30), we further divide (P2-a) into two subproblems: 1) maximizing UAV received power, i.e., $P_{i,j}10^{-\left(L_{i,j}^{\mathrm{D2V}}/10\right)}$ and 2) minimizing GD interference power, i.e., $\sum_{k\in\mathcal{S}_j\setminus i} P_{k,j}10^{-\left(L_{k,j}^{\mathrm{D2V}}/10\right)}$. It is easy to optimize variable *P* by Proposition 2, so we focus on optimizing the ground-to-air path loss in (3). By introducing the optimal transmit power into (30), we redefine (P2-a) as

$$\min_{h_{i,j}} \quad 10\log_{10}\frac{\left(4\pi h_{i,j}\right)^2}{G_l\lambda^2}+\frac{\left(\eta_{\mathrm{LoS}}-\eta_{\mathrm{NLoS}}\right)}{1+a\exp\left[-b\left(\pi/2-a\right)\right]}+\eta_{\mathrm{NLoS}}, \quad (31)$$

where $d_{i,j}=h_{i,j}$ holds iff $\mathbb{P}_{\mathrm{LoS}}^{i,j}=1$. Thus, the minimum of (31) is obtained when UAV *j* hovers over TDCC *i*, and the boundary of $h_{i,j}$ is given by the following proposition.

*Proposition 3:* The lower and upper bounds of $h_{i,j}$, denoted by $h_{\min}$ and $h_{\max}$, are given by

$$\begin{cases} h_{\min}=\sqrt{d_0 G_l\lambda^2\left(\ln\left(d_0\left(4\pi\right)^2/G_l\lambda^2\right)+1\right)\left(4\pi\right)^{-2}}, \\ h_{\max}=\sqrt{P_{\max}G_l\lambda^2 10^{-(C/10)}\left[\left(4\pi\right)^2\gamma\left(N_0+\sum_{k\in\mathcal{S}_j\setminus i}P_0\right)\right]^{-1}}, \end{cases} \quad (32)$$

where $C=\left(\eta_{\mathrm{LoS}}-\eta_{\mathrm{NLoS}}\right)\left\{1+a\exp\left[-b\left(\pi/2-a\right)\right]\right\}^{-1}+\eta_{\mathrm{NLoS}}$.

*Proof:* See Appendix A.

Using Proposition 3, we obtain a set of heights, denoted by $\left\{h_{i,j}\right\}$. Furthermore, the optimal height of UAV *j* is given by $h_{i,j}^*=\arg\min_{h_{i,j}}\left\{h_{i,j}\right\}$, and its 2-D coordinate is solved by

$$\min_{x_j^{\mathrm{UAV}},y_j^{\mathrm{UAV}}} \quad \sum_{i\in\mathcal{S}_j}\left(x_j^{\mathrm{UAV}}-x_i\right)^2+\left(y_j^{\mathrm{UAV}}-y_i\right)^2, \quad (33)$$

which yields $\mathbf{u}_j^*=\left(x_j^{\mathrm{UAV}*},y_j^{\mathrm{UAV}*},h_{i,j}^*+z_i^*\right)$, where $z_i^*$ is the DEM point's *z*-coordinate that is closest to UAV *j*.

Now, given $\left\{\mathbf{u}_j^*\right\}$, we aim to minimize $t_m$ by optimal control, which leads to the following motion planning problem

(P2-b): $$\min_t \quad t=t_m \quad (34)$$

$$\text{s.t.} \quad \begin{bmatrix}\dot{\mathbf{p}}(t)\\ \ddot{\mathbf{p}}(t)\end{bmatrix}=\begin{bmatrix}0 & 1\\ 0 & 0\end{bmatrix}\begin{bmatrix}\mathbf{p}(t)\\ \dot{\mathbf{p}}(t)\end{bmatrix}+\begin{bmatrix}0\\ 1\end{bmatrix}\alpha(t), \quad (35)$$

where $\mathbf{p}(t)=\left[x_j^{\mathrm{UAV}},y_j^{\mathrm{UAV}},z_j^{\mathrm{UAV}}\right]$, $\alpha(t)=\begin{cases}\mathbf{a}_{\max}, & 0\le t\le\overline{t}\\ \mathbf{a}_{\min}, & \overline{t}\le t\le t^*\end{cases}$, and $t^*$ and $\overline{t}$ are the shortest flight time and switching time, respectively.

*Proposition 4:* $t^*$ and $\overline{t}$ are given by

$$\begin{cases} t^*=\sqrt{\dfrac{4d\left(t^*\right)g\cos\left(\theta\right)}{\left\|\mathbf{a}_{\max}\right\|^2+\left\|\mathbf{a}_{\max}\right\|2g\cos\left(\theta\right)}}, \\ \overline{t}=\dfrac{t^*\left\|\mathbf{a}_{\max}\right\|+2g\cos\left(\theta\right)}{2g\cos\left(\theta\right)}, \end{cases} \quad (36)$$

where $d\left(t^*\right)$ is the displacement, $\theta$ is the pitch angle, and $\mathbf{a}_{\max}$ is the maximum acceleration.

*Proof:* See Appendix B.

Following (36), four proportional-differential controllers are employed to generate expected commands, as shown in Fig. 5. The small-angle control model is given by [28]

$$\begin{cases} \ddot{x}(t)=\dfrac{f}{m}\left(\sin(\psi)\sin(\phi)+\cos(\psi)\sin(\theta)\cos(\phi)\right)+\dfrac{f_x}{m}, \\ \ddot{y}(t)=\dfrac{f}{m}\left(-\cos(\psi)\sin(\varphi)+\cos(\phi)\sin(\theta)\sin(\psi)\right)+\dfrac{f_y}{m}, \\ \ddot{z}(t)=\dfrac{f}{m}\cos(\theta)\cos(\phi)-g+\dfrac{f_z}{m}, \\ \ddot{\phi}(t)=\dfrac{\tau_x}{I_{xx}},\ddot{\theta}(t)=\dfrac{\tau_y}{I_{yy}},\ddot{\psi}(t)=\dfrac{\tau_z}{I_{zz}}, \end{cases} \quad (37)$$

where $\phi$, $\theta$, and $\psi$ are roll, pitch, and yaw angles, respectively, $f$ is the total lift, $\mathbf{F}=\mathbf{R}\cdot\mathbf{f}=\left[f_x,f_y,f_z\right]$ is the external force, and $\boldsymbol{\tau}=\left[\tau_x,\tau_y,\tau_z\right]$ is the torque vector, which meets

$$\begin{bmatrix}f\\ \tau_x\\ \tau_y\\ \tau_z\end{bmatrix}=\underbrace{\begin{bmatrix}c_t & c_t & c_t & c_t\\ 0 & -r_f c_t & 0 & r_f c_t\\ r_f c_t & 0 & -r_f c_t & 0\\ c_m & -c_m & c_m & -c_m\end{bmatrix}}_{\mathbf{N}}\underbrace{\begin{bmatrix}\varpi_1^2\\ \varpi_2^2\\ \varpi_3^2\\ \varpi_4^2\end{bmatrix}}_{\mathbf{n}}, \quad (38)$$

where $r_f$ is the fuselage radius, and $c_t$ and $c_m$ are the thrust and torque coefficients, respectively. Using (38), the position controller and attitude controller are defined as follows [29]

$$\begin{cases} \Theta=g^{-1}\mathbf{A}^{-1}\left(k_p\Delta\left(v_h\right)+k_d\dot{\Delta}\left(v_h\right)-\mathbf{m}\right), \\ f=m\left(g+k_p\Delta\left(v_v\right)+k_d\dot{\Delta}\left(v_v\right)\right)-f_z, \end{cases} \quad (39)$$

$$\begin{cases} \omega=\dot{\Theta}, \\ \boldsymbol{\tau}=k_p\Delta\left(\omega\right)+k_d\dot{\Delta}\left(\omega\right), \end{cases} \quad (40)$$

where $\Theta=\left[\phi,\theta\right]^T$, $\mathbf{A}^{-1}=\left(\begin{bmatrix}\cos(\psi) & -\sin(\psi)\\ \sin(\psi) & \cos(\psi)\end{bmatrix}\begin{bmatrix}0 & 1\\ -1 & 0\end{bmatrix}\right)^{-1}$, $\mathbf{m}=\left[f_x m^{-1},f_y m^{-1}\right]^T$, $k_p$ and $k_d$ are coefficients, and $\Delta(\cdot)$ denotes the error between the current and expected states.

Using (38)-(40), the allocation controller yields the expected motor speeds

$$\mathbf{n}=\mathbf{N}^{-1}\left[f,\tau_x,\tau_y,\tau_z\right]^T, \quad (41)$$

where $\mathbf{n}$ is input into the power controller to obtain the expected command, denoted by $\Phi=k_p\mathbf{n}+k_d\dot{\mathbf{n}}$.

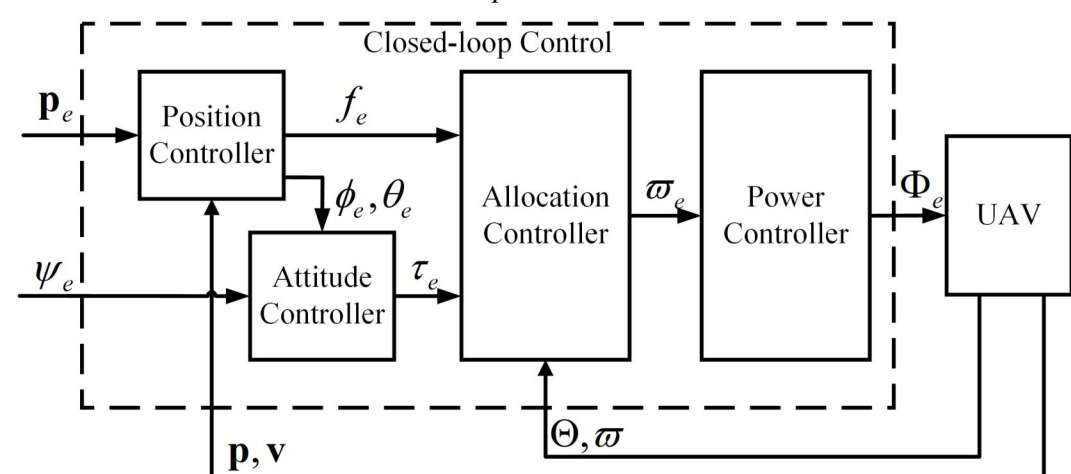

**Fig. 5.** Diagram of closed-loop control. The subscript "*e*" denotes expectation.

*2) Optimal Network Configurations:* By solving (P2-a) and (P2-b), we obtain the optimal path of an individual UAV. Furthermore, two network configurations, including $\mathcal{M}$ and $P_{i,j}$, are optimized to achieve the best time-energy tradeoff

$$\min_{x,\mathbf{y}} \quad \max\left\{\left\{T_j\right\},\left\{T_i\right\}\right\} \quad (42)$$

$$\min_{x,\mathbf{y}} \quad \sum_{j\in\mathcal{M}} E_j + \sum_{i\in\mathcal{T}} E_i \tag{43}$$

$$\text{s.t.} \quad 1 \le x \le |\mathcal{T}|, P_0 10^{\left(L_{i,j}^{\mathrm{D2V}}/10\right)} \le y_i \le P_{\max}, \forall y_i \in \mathbf{y}, \tag{44}$$

$$E_j\left(x,\mathbf{y}\right) \le E_{\max}, (13)-(16), \forall j \in \mathcal{M}, \tag{45}$$

where $x$ and $\mathbf{y}$ denote the number of UAVs and the vector of transmit power, respectively.

Then, the MOEA/D is employed to simplify (42)-(45) as [30]

$$\min_{x,\mathbf{y}} \quad \mathcal{F}\left(\mathbf{x}|\boldsymbol{\lambda},\mathbf{z}^*\right) = \max\left\{\lambda_i \left|\mathcal{G}_i\left(x,\mathbf{y}\right) - z_i^*\right|\right\} \tag{46}$$

$$\text{s.t.} \quad 1 \le i \le 2, (44)-(45), \tag{47}$$

where $\mathbf{x} = [x,\mathbf{y}]$, $\boldsymbol{\lambda} = [\lambda_1, \lambda_2]$ is the weight vector, $\mathbf{z}^* = [z_1^*, z_2^*]$ is the reference vector, and $\mathcal{G}_i$ denotes objective function $i$.

To solve (46), we minimize the interval between $\mathcal{F}\left(\mathbf{x}|\boldsymbol{\lambda},\mathbf{z}^*\right)$ and $\mathcal{F}\left(\mathbf{x}|\boldsymbol{\lambda}',\mathbf{z}^*\right)$ and have

$$\mathcal{E} = \left\{\left\{\mathbf{x}_i\right\}, \mathcal{F}\left(\mathbf{x}_i\right), \left\{z_i^*\right\}, \left\{\mathcal{F}^*\left(\mathbf{x}_i\right)\right\}\right\}, \tag{48}$$

where $\mathbf{x}_i$ is the current solution to objective function $i$, $\mathcal{F}\left(\mathbf{x}_i\right)$ denotes the function value, $z_i^*$ is the optimal value of objective function $i$, and $\left\{\mathcal{F}^*\left(\mathbf{x}_i\right)\right\}$ is the PF.

Since any pair of solutions in the PF cannot dominate each other, to determine the best tradeoff point, we select EE as our decision-making indicator, which is defined as the reciprocal of the weighted sum of (42) and (43)

$$\max_{\{T_j\},\{T_i\},\{E_j\},\{E_i\}} \quad \left[\Gamma\mathbb{N}\left(\max\left\{\{T_j\},\{T_i\}\right\}\right) + (1-\Gamma)\mathbb{N}\left(\sum_{j\in\mathcal{M}} E_j + \sum_{i\in\mathcal{T}} E_i\right)\right]^{-1}, \tag{49}$$

where $\Gamma$ is the priority weight. In this article, we set $\Gamma = 0.5$, indicating that shortening ERT is as important as reducing EC.

## IV. Simulation Results and Discussions

In our simulations, $\mathbf{G}$ is located in a 9 km$^2$ geographic area, which is derived from the ASTER Global DEM available at https://doi.org/10.5067/ASTER/AST_L1A.003, and the size of $\mathbf{G}$ belongs to {50, 100, 150, 200}. Table II summarizes the simulation parameters. To validate our proposed approach, we compare the following benchmark schemes: 1) Only D2D [6]: directly collect data through D2D channels; 2) Static D2V [7]: dispatch UAVs to collect data, but the locations of UAVs are fixed; 3) Dynamic D2V [10]: different from the static scheme, UAVs continuously update locations to receive data; 4) Hybrid scheme [11]: UAVs work as static aerial BSs and mobile relays. In contrast, we propose a hybrid D2D and D2V network, and the advantages of our proposed approach include: 1) The D2D network is established before UAVs reach hovering locations, which helps to shorten ERT and reduce EC; 2) The number of UAVs is adaptively determined, which is more realistic; 3) The practical performance of UAVs is taken into account, which contributes to better energy management; 4) EE is selected to characterize the conflict between ERT and EC.

Table II: Simulation Parameters

| Parameter | Description | Value |
| --- | --- | --- |
| $a$ | Carrier frequency constant | 11.95 |
| $b$ | Environmental constant | 0.14 |
| $\eta_{\mathrm{LoS}}$ | Additional LoS path loss | 3.0 dB |
| $\eta_{\mathrm{NLoS}}$ | Additional NLoS path loss | 23.0 dB |
| $\mu$ | Packet size | 1024.0 byte |
| $G_l$ | Antenna Gain | 1.0 |
| $N_0$ | Noise power | -130.0 dBm |
| $\sigma$ | Standard deviation | 3.65 dB |
| $\varepsilon$ | Outage probability threshold | 0.01 |
| $P_{\max}$ | Maximum transmit power | 10.0 mW |
| $C_{\max}$ | Capacity threshold | 30 |
| $m$ | Mass | 4.0 kg |
| $g$ | Gravity acceleration | 9.8 m/s$^2$ |
| $\rho$ | Air density | 1.225 kg/m$^3$ |
| $c_d$ | Drag coefficient | 0.117 N/(m/s)$^2$ |
| $c_b$ | Blade chord | 0.1 m |
| $N_b$ | Number of blades | 4 |
| $r_d$ | Rotor disk radius | 0.5 m |
| $r_f$ | Fuselage radius | 0.3 m |
| $k_p$ | Proportional coefficient | 0.2 |
| $k_d$ | Differential coefficient | 0.001 |
| $I_{xx}$ | $x$-axis inertia moment | 6.302×10$^{-2}$ kg·m$^2$ |
| $I_{yy}$ | $y$-axis inertia moment | 6.302×10$^{-2}$ kg·m$^2$ |
| $I_{zz}$ | $z$-axis inertia moment | 1.171×10$^{-2}$ kg·m$^2$ |
| $f_{\max}$ | Maximum lift | 68.1 N |
| $c_t$ | Thrust coefficient | 2.646×10$^{-5}$ N/(rad/s)$^2$ |
| $c_m$ | Torque coefficient | 4.411×10$^{-7}$ N·m/(rad/s)$^2$ |
| $P_{\mathrm{hover}}$ | Hovering power | 99.4 W |

Note that the simulation results in this section are averaged over a large number of independent experiments, implemented using Python running on a computer with an Intel Core i7-2.6 GHz CPU and 64 GB memory, and the primarily used packages include CVXPY, NETWORKX, and PYMOO. CVXPY is used to solve (P1-b) and (P2-a), NETWORKX establishes the D2D network by solving (P1-a), and (P2-b) is optimized by PYMOO. The simulation steps are shown in Algorithm 4, where steps 4-5 and steps 6-13 are used to identify TDCCs and schedule UAVs, respectively. During each iteration, the Hungarian method [25] is employed to assign missions, and the optimal path planning is implemented by steps 9-10. The computational complexity of Algorithm 4 is $\mathcal{O}\left(|\mathcal{C}||\mathbf{v}_1|N^3 + T|\mathcal{Z}||\mathcal{M}|^4|\mathcal{T}|^3\right)$, where $|\mathcal{C}| = |\mathcal{T}|$ and $\mathcal{Z}$ is the population size of the MOEA/D.

**Algorithm 4** Algorithm for System Simulation

1: **Input:** Set of GDs $\mathcal{N}$, set of DEM points $\mathcal{D}$, and number of iterations $T$
2: **Output:** Optimal location of UAVs $\{\mathbf{u}_j^*\}$
3: **Initialize:** Initial configuration $\leftarrow \{\mathcal{M}_0, \{\mathbf{u}_j^0\}, \{P_{i,j}^0\}\}$
4: Generate $|\mathcal{T}|$ clusters by **Algorithm 3**
5: Identify TDCCs from $\mathcal{T}$ using the TOPSIS model
6: **for** $t = 1,...,T$ **do**
7: Generate $\{\mathcal{S}_j\}$ using the Hungarian method
8: **for** $j \in \mathcal{M}_t$ **do**
9: Update $\{h_{i,j}\}, \forall i \in \mathcal{S}_j$, by Proposition 3
10: Update $\mathbf{u}_j^t$ and generate 3-D paths by Proposition 4
11: **end for**
12: Update $\mathcal{M}_t$ and $\{P_{i,j}^t\}$ by the MOEA/D
13: **end for**
14: return $\{\mathbf{u}_j^t\}$

Fig. 6 shows ONP derived from Algorithm 3 and the other three schemes: 1) clique partitioning (CP) with undirected and unweighted edges 2) CP with directed and unweighted edges and 3) CP with undirected and weighted edges. Figs. 6(a)-6(d) demonstrate the convergence of $Q$ with different network sizes, where $Q$ converges to 1.1, 1.3, 1.4, and 2.6 in the four scenarios, respectively. It is shown that Algorithm 3 gives a higher $Q$ than benchmark schemes. Since Algorithm 3 iteratively optimizes adjacent matrix $\mathbf{L}$ and power matrix $\mathbf{P}$, D2D connectivity is remarkably strengthened. Thus, a GD in each cluster transmits its collected emergency data to its corresponding TDCC. Note that if a GD is located in a remote area or blocked by terrain, it will be identified as a TDCC and served by a UAV.

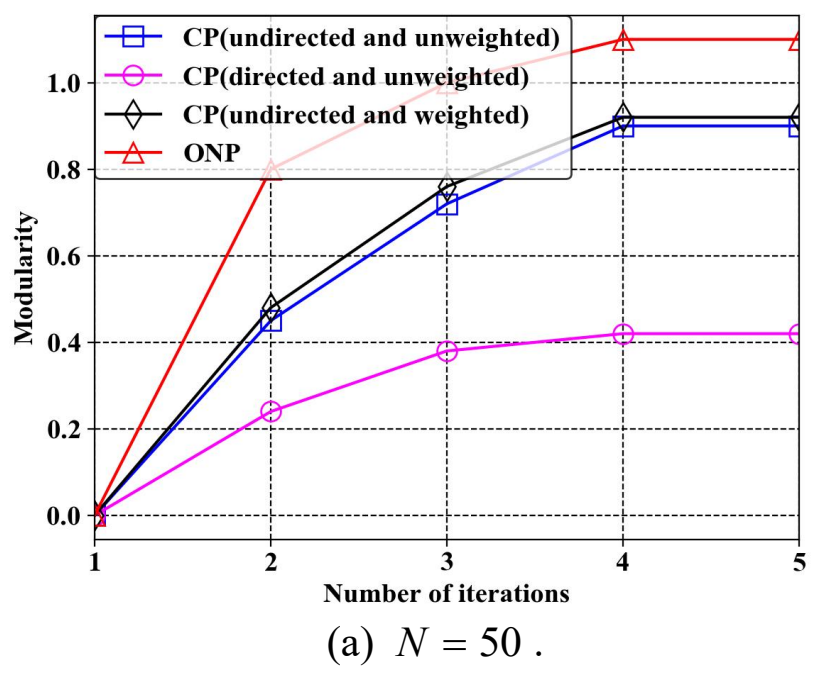

(a) $N = 50$ .

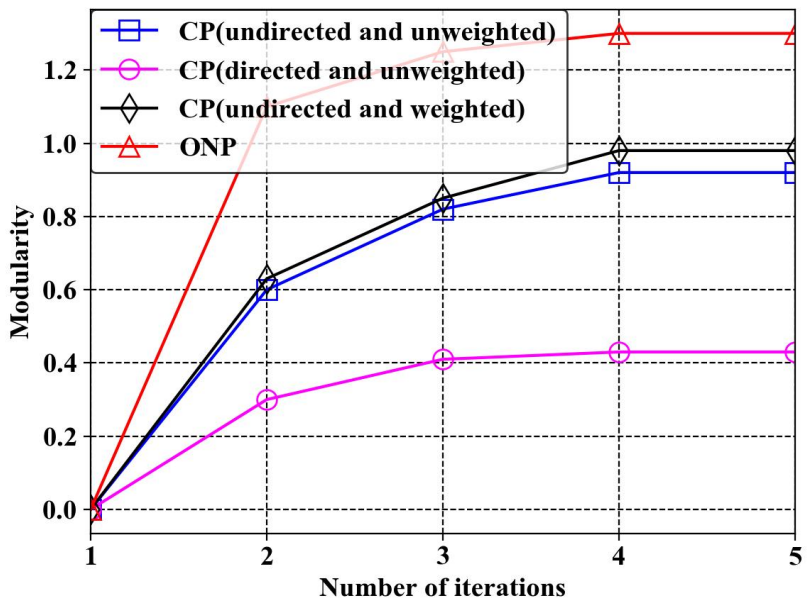

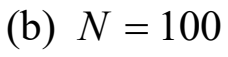

(b) $N = 100$ .

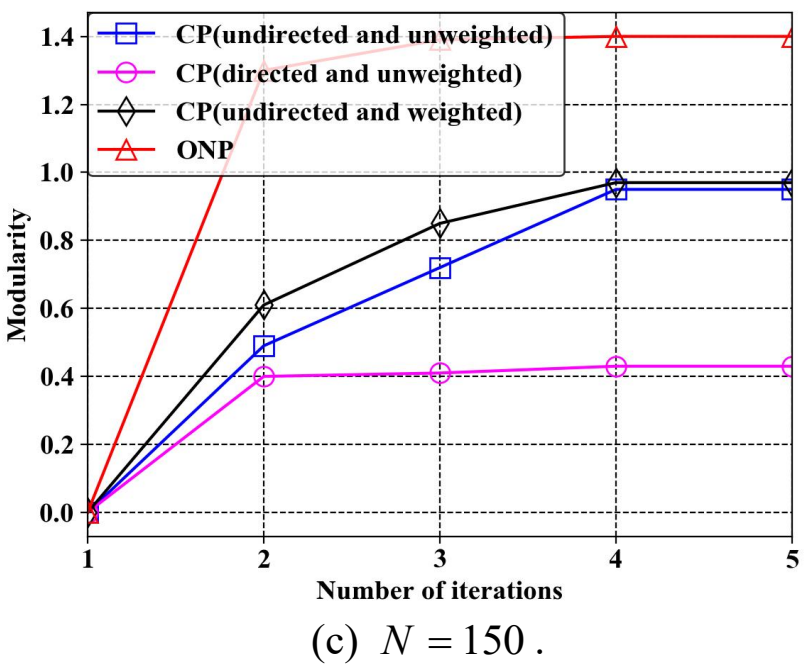

(c) $N = 150$ .

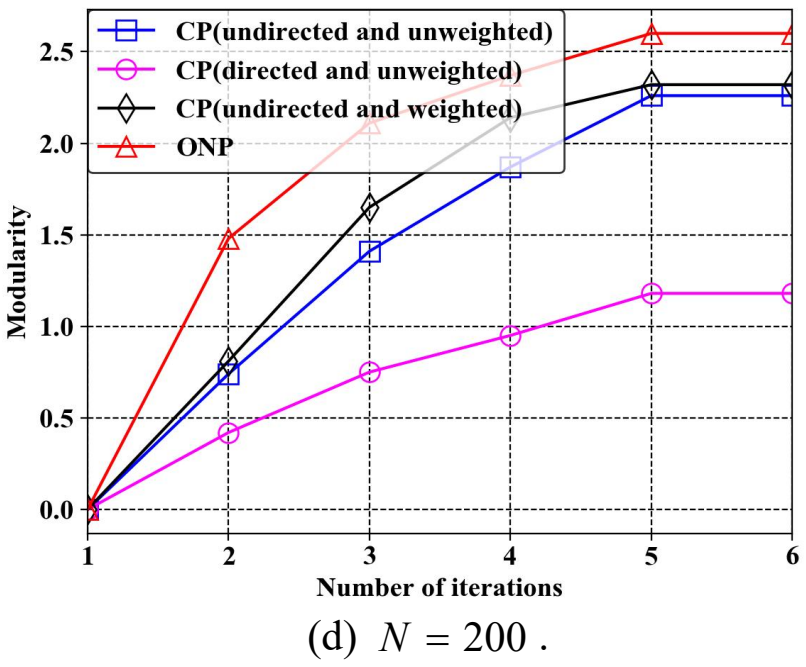

(d) $N = 200$ .

**Fig. 6.** Convergence of modularity.

Fig. 7 shows the performance of optimal path planning. By waypoint and motion planning, UAVs dynamically update their locations to maximize capacity and minimize movement time. First, Fig. 7(a) shows that the average path loss converges to 58.93 dB, indicating that UAVs reach their optimal hovering waypoints to maximize data transmission rates in the uplink. Second, Fig. 7(b) shows that the average ground-to-air latency decreases from 11.04 s to 4.20 s. Furthermore, Fig. 7(c) proves that the optimal control (OC) algorithm saves 21.60% of flight time compared to a uniform-velocity (UV) scheme. Note that UAVs must hover over TDCCs for a while to receive complete data. Thus, the initial and final velocities between two adjacent waypoints are 0 m/s. Finally, as shown in Fig. 7(d), the average motion EC reaches 103.83 kJ, which suggests that the required number of UAVs is mainly constrained by onboard energy.

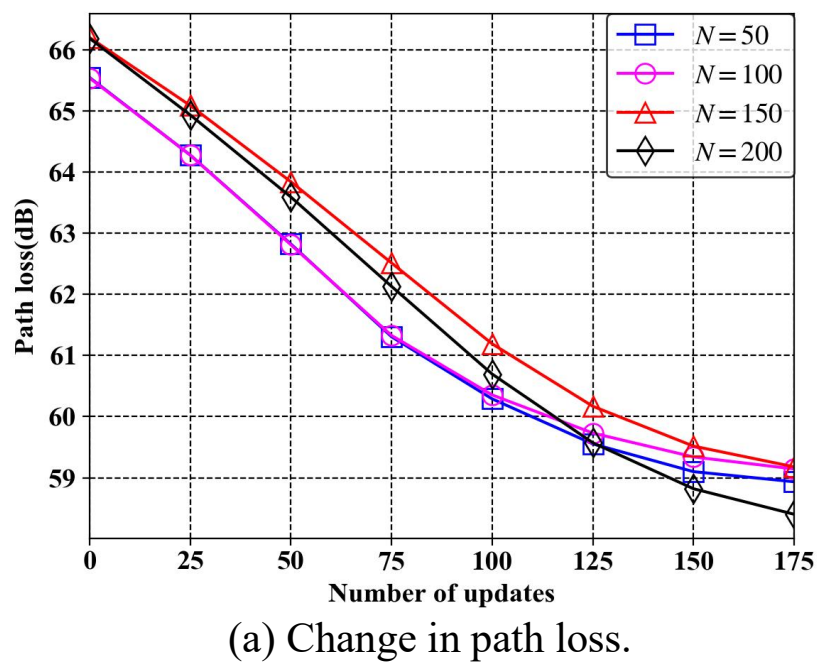

(a) Change in path loss.

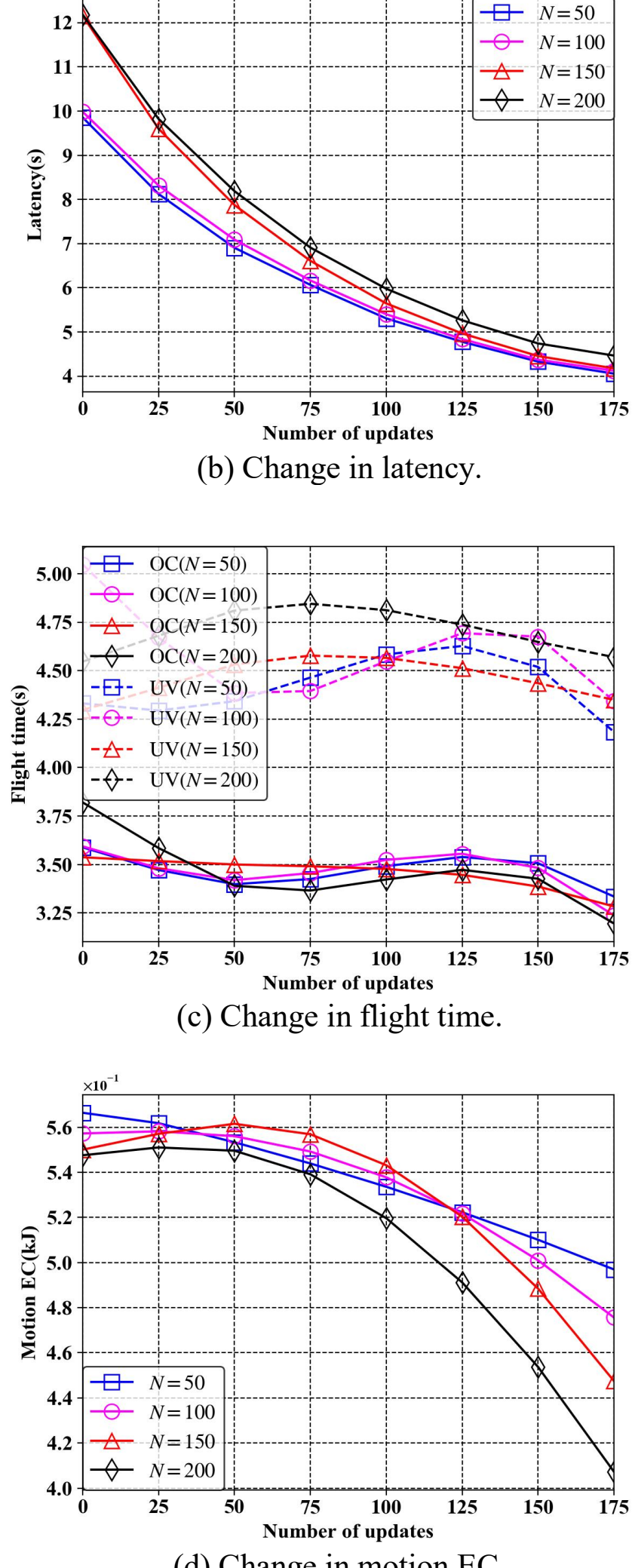

(b) Change in latency.

(c) Change in flight time.

(d) Change in motion EC.

**Fig. 7.** Optimal path planning.

Fig. 8 shows the results of multiobjective optimization. It is clear that increasing the number of UAVs reduces ERT, which causes more EC, indicating the contradiction between (44) and (45). To simultaneously minimize the maximum ERT and the total EC, we employ the MOEA/D to optimize the number of UAVs and the transmit power of TDCCs. Fig. 8(a) plots the PFs in the four scenarios, and the corresponding solutions are presented in Fig. 8(b), where we normalize the values of ERT and EC to fall between 0 and 1 and set the priority weight $\Gamma$ to 0.5. In this article, EE serves as the performance indicator, and the optimal solution is achieved when EE reaches its maximum. As shown in Fig. 8(c), the optimal EE is 4.85, 4.75, 4.51, and 4.18, respectively. Note that the variation in EE is dependent on $\Gamma$. In Fig. 8(d), we compare EE with different priority weights. It is observed that EE increases with $\Gamma$, suggesting that the increment in EC is greater than that of ERT with an increase in the number of UAVs. Consequently, we compare our proposed approach with the other four benchmark schemes. As shown in Fig. 8(e), our proposed approach improves EE by 7%, 12%, 20%, and 23%, respectively.

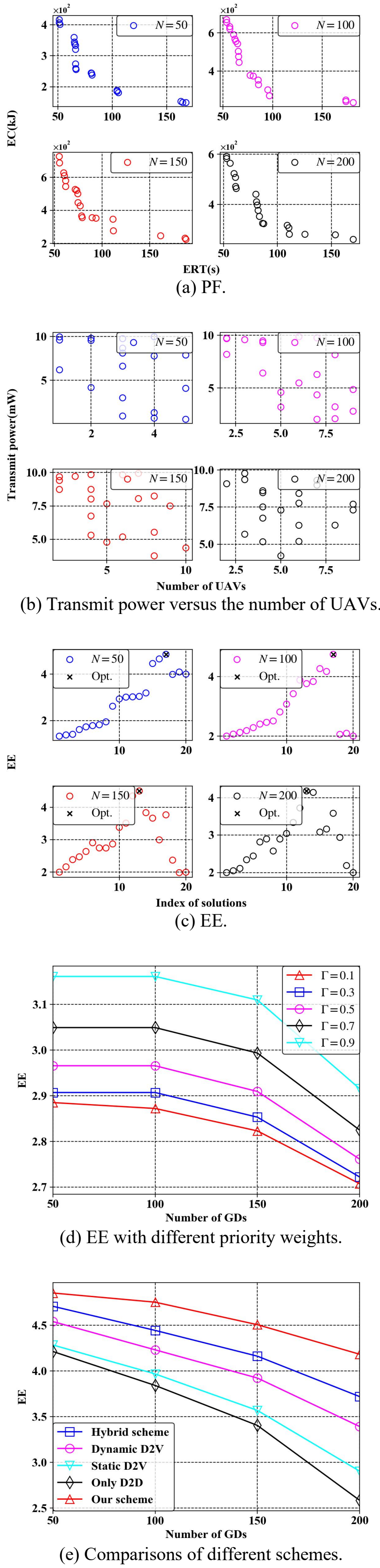

(a) PF.

(b) Transmit power versus the number of UAVs.

(c) EE.

(d) EE with different priority weights.

(e) Comparisons of different schemes.

**Fig. 8.** Multiobjective optimization.

## V. CONCLUSION

This article newly proposed a hybrid D2D and D2V network for collecting and transmitting emergency data. First, the initial D2D network was divided into clusters, and TDCCs in clusters cached emergency data from other members. Second, given the locations of TDCCs, we updated the locations and attitudes of UAVs to shorten the time for data transmission and movement. Consequently, the best time-energy tradeoff was achieved by maximizing EE, contributing to implementing energy-efficient emergency communications.

However, there are other limitations not addressed yet, some of which would motivate our future work: 1) In our simulations, we assume that environmental information (EI) is available so that the EOC can plan real-time paths for UAVs. In practice, EI is variable, so the EOC should monitor EI in real time; 2) In the D2D network, we calculate all SPs for each pair of hosts, which increases system EC, so energy-aware routing protocols can be applied to increase the lifetime of D2D networks; 3) Compared to a quadcopter, a fixed-wing UAV can provide longer cruising time but cannot hover at a fixed position. Therefore, it will be promising to jointly deploy rotary-wing and fixed-wing UAVs. Rotary-wing UAVs serve as aerial BSs hovering over densely distributed GDs, and fixed-wing UAVs are employed to cover a larger area.

## APPENDIX

### A. Proof of Proposition 3

By introducing (25) into (30), we have

$$\begin{aligned}\frac{\left(4\pi h_{i,j}\right)^2}{G_l\lambda^2} &\leq P_{i,j}10^{-(C/10)}\left[\gamma\left(N_0+\sum_{k\in\mathcal{S}_j\setminus i}P_{k,j}10^{-\left(L_{k,j}^{\mathrm{D2V}}/10\right)}\right)\right]^{-1}\\ &\leq P_{\max}10^{-(C/10)}\left[\gamma\left(N_0+\sum_{k\in\mathcal{S}_j\setminus i}P_0\right)\right]^{-1}.\end{aligned} \tag{50}$$

Using (50), the upper bound of $h_{i,j}$ satisfies

$$h_{i,j}\leq h_{\max}=\sqrt{P_{\max}G_l\lambda^2 10^{-(C/10)}\left[\left(4\pi\right)^2\gamma\left(N_0+\sum_{k\in\mathcal{S}_j\setminus i}P_0\right)\right]^{-1}}. \tag{51}$$

Furthermore, we rewrite (31) as

$$\begin{aligned}10\log_{10}\frac{\left(4\pi h_{i,j}\right)^2}{G_l\lambda^2}&=\frac{10}{\ln 10}\ln\left(\frac{\left(4\pi h_{i,j}\right)^2}{G_l\lambda^2}\right),\\ \ln\left(\frac{\left(4\pi h_{i,j}\right)^2}{G_l\lambda^2}\right)&=\ln\left[d_0\left(\frac{\left(4\pi h_{i,j}\right)^2}{d_0G_l\lambda^2}-1+1\right)\right]\\ &\simeq\ln\left(d_0\right)+\sum_{n=1}^{\infty}\frac{(-1)^{n-1}}{n}\left(\frac{\left(4\pi h_{i,j}\right)^2}{d_0G_l\lambda^2}-1\right)^n,\end{aligned} \tag{52}$$

where $d_0$ is derived from the positioning error ( $\mu_0=20.0$ and $\sigma_0^2=5.0$ ). Following (52), we closely approximate the lower bound of $h_{i,j}$ using the Taylor's expansion at $n=1$ [31]

$$\begin{aligned}&\ln\left(d_0\right)+\sum_{n=1}^{\infty}\frac{(-1)^{n-1}}{n}\left(\frac{\left(4\pi h_{i,j}\right)^2}{d_0G_l\lambda^2}-1\right)^n\overset{n=1}{=}\ln\left(d_0\right)+\left(\frac{\left(4\pi h_{i,j}\right)^2}{d_0G_l\lambda^2}-1\right),\\ &\ln\left(d_0\right)+\left(\frac{\left(4\pi h_{i,j}\right)^2}{d_0G_l\lambda^2}-1\right)=\ln\left(\frac{\left(4\pi h_{i,j}\right)^2}{G_l\lambda^2}\right)\overset{h_{i,j}\geq d_0}{\geq}\ln\left(\frac{\left(4\pi d_0\right)^2}{G_l\lambda^2}\right),\\ &h_{i,j}\geq h_{\min}=\sqrt{d_0G_l\lambda^2\left(\ln\left(d_0\left(4\pi\right)^2/G_l\lambda^2\right)+1\right)\left(4\pi\right)^{-2}}.\end{aligned} \tag{53}$$

Clearly, (51) and (53) prove the proposition.

### B. Proof of Proposition 4

The velocity constraint is expressed as

$$\left\|\mathbf{a}_{\max}\right\|\bar{t}+\left\|\mathbf{a}_{\min}\right\|\left(t^*-\bar{t}\right)=0, \tag{54}$$

which yields the expression of $\bar{t}$

$$\bar{t}=\frac{-t^*\left\|\mathbf{a}_{\min}\right\|}{\left\|\mathbf{a}_{\max}\right\|-\left\|\mathbf{a}_{\min}\right\|}. \tag{55}$$

**Fig. 9.** Force diagram of a UAV. $\mathbf{z}_b$ is the unit vector in the body frame.

Given $\mathbf{u}_a=\left(x_a^{\mathrm{UAV}},y_a^{\mathrm{UAV}},z_a^{\mathrm{UAV}}\right)$ and $\mathbf{u}_b=\left(x_b^{\mathrm{UAV}},y_b^{\mathrm{UAV}},z_b^{\mathrm{UAV}}\right)$, we need to guarantee that the resultant force of a UAV is in the same direction as $\overrightarrow{\mathbf{u}_a\mathbf{u}_b}$, as shown in Fig. 9, which yields

$$p=\arcsin\left(\frac{\left\|\mathbf{F}\right\|\sin\left(q\right)+mg\sin\left(\theta\right)}{\left\|\mathbf{f}_{\max}\right\|}\right), \tag{56}$$

where $p$ is the angle between $\overrightarrow{\mathbf{u}_a\mathbf{u}_b}$ and $\mathbf{f}_{\max}$, $\mathbf{f}_{\max}$ is the maximum lift, $q=\arccos\left(\frac{\mathbf{F}\cdot\overrightarrow{\mathbf{u}_a\mathbf{u}_b}}{\left\|\mathbf{F}\right\|\cdot\left\|\overrightarrow{\mathbf{u}_a\mathbf{u}_b}\right\|}\right)$, and $\theta$ and $\psi$ satisfy

$$\begin{cases}F\cos\left(\theta_1\right)=\left\|\mathbf{F}\right\|\cos\left(\theta_2\right)+\left\|\mathbf{f}_{\max}\right\|\cos\left(\theta\right),\\ \left\|\mathbf{F}\right\|\sin\left(\theta_2\right)\sin\left(\psi_1-\psi_2\right)=\left\|\mathbf{f}_{\max}\right\|\sin\left(\theta\right)\sin\left(\psi-\psi_1\right),\end{cases} \tag{57}$$

where $\theta_1=\arccos\left(\frac{z_b^{\mathrm{UAV}}-z_b^{\mathrm{UAV}}}{d\left(t^*\right)}\right)$, $\theta_2=\arccos\left(\frac{f_z}{\left\|\mathbf{F}\right\|}\right)$,

$\psi_1=\arctan\left(\frac{y_b^{\mathrm{UAV}}-y_a^{\mathrm{UAV}}}{x_b^{\mathrm{UAV}}-x_a^{\mathrm{UAV}}}\right)$, $\psi_2=\arctan\left(\frac{f_y}{f_x}\right)$, and

$$F=\sqrt{\left\|\mathbf{f}_{\max}\right\|^2+\left\|\mathbf{F}\right\|^2+2\left\|\mathbf{f}_{\max}\right\|\left\|\mathbf{F}\right\|\cos\left(p+q\right)}-mg\cos\left(\theta\right).$$

By projecting $\mathbf{f}_{\max}$ and $\mathbf{F}$ on $\overrightarrow{\mathbf{u}_a\mathbf{u}_b}$, we have

$$\begin{cases}\left\|\mathbf{a}_{\max}\right\|=\dfrac{\left\|\mathbf{f}_{\max}\right\|\sin\left(\theta\right)+\left\|\mathbf{F}\right\|\cos\left(q\right)-mg\cos\left(\theta\right)}{m}.\\ \left\|\mathbf{a}_{\min}\right\|=\left\|\mathbf{a}_{\max}\right\|+2g\cos\left(\theta\right).\end{cases} \tag{58}$$

Using (55) and (58), $t^*$ is solved by

$$
\begin{aligned}
d\left(t^*\right) &= \frac{1}{2}\left\|\mathbf{a}_{\max}\right\|\bar{t}^2 + \left\|\mathbf{a}_{\max}\right\|\bar{t}\left(t^*-\bar{t}\right) + \frac{1}{2}\left\|\mathbf{a}_{\min}\right\|\left(t^*-\bar{t}\right)^2, \\
d\left(t^*\right) &\overset{(55)}{=} \frac{1}{2}\left\|\mathbf{a}_{\max}\right\|\left(\frac{-t^*\left\|\mathbf{a}_{\min}\right\|}{\left\|\mathbf{a}_{\max}\right\|-\left\|\mathbf{a}_{\min}\right\|}\right)^2 + \frac{1}{2}\left\|\mathbf{a}_{\min}\right\|\left(t^* - \frac{-t^*\left\|\mathbf{a}_{\min}\right\|}{\left\|\mathbf{a}_{\max}\right\|-\left\|\mathbf{a}_{\min}\right\|}\right)^2 \\
&\quad + \left\|\mathbf{a}_{\max}\right\|\left(\frac{-t^*\left\|\mathbf{a}_{\min}\right\|}{\left\|\mathbf{a}_{\max}\right\|-\left\|\mathbf{a}_{\min}\right\|}\right)\left(t^* + \frac{t^*\left\|\mathbf{a}_{\min}\right\|}{\left\|\mathbf{a}_{\max}\right\|-\left\|\mathbf{a}_{\min}\right\|}\right), \\
d\left(t^*\right) &\overset{(58)}{=} \frac{\left\|\mathbf{a}_{\max}\right\|\left(t^*\left(\left\|\mathbf{a}_{\max}\right\|+2g\cos\left(\theta\right)\right)\right)^2}{8g^2\cos\left(\theta\right)^2} \\
&\quad - \frac{t^{*2}\left\|\mathbf{a}_{\max}\right\|^2\left(\left\|\mathbf{a}_{\max}\right\|+2g\cos\left(\theta\right)\right)}{4g^2\cos\left(\theta\right)^2} \\
&\quad + \frac{t^{*2}\left\|\mathbf{a}_{\max}\right\|^2\left(\left\|\mathbf{a}_{\max}\right\|+2g\cos\left(\theta\right)\right)}{8g^2\cos\left(\theta\right)^2}, \\
t^* &= \sqrt{\frac{4d\left(t^*\right)g\cos\left(\theta\right)}{\left\|\mathbf{a}_{\max}\right\|^2+2\left\|\mathbf{a}_{\max}\right\|g\cos\left(\theta\right)}}.
\end{aligned}
\tag{59}
$$

Clearly, (55), (58), and (59) prove the proposition.

## ACKNOWLEDGMENT

The author thanks Dr. Chongcheng Chen for proofreading and diagram conceptualization.